\documentclass{nature}

\title{Evidence of a dynamically evolving Galactic warp}

\usepackage{graphicx}	
\usepackage{amsmath}	
\usepackage{amssymb}	

\usepackage{lineno}

\usepackage{hyperref}
\hypersetup{ colorlinks, citecolor=blue, filecolor=black, linkcolor=blue, urlcolor=blue }

\newcommand{\gaia}{\textit{Gaia}}
\newcommand{\gdrtwo}{\textit{Gaia}~DR2}
\newcommand{\kmskpc}{$\text{\, km} \text{\, s}^{-1} \text{\, kpc}^{-1}$}

\author{E.~Poggio$^{1}$,
R.~Drimmel$^{1}$,
R.~Andrae$^{2}$,
C.~A.~L.~Bailer-Jones$^{2}$,
M.~Fouesneau$^{2}$,
M.~G.~Lattanzi$^{1}$,
R.~L.~Smart$^{1}$, \&
A.~Spagna$^{1}$ 
}

\makeatletter
\let\saved@includegraphics\includegraphics
\AtBeginDocument{\let\includegraphics\saved@includegraphics}
\renewenvironment*{figure}{\@float{figure}}{\end@float}
\makeatother

\begin{document}


\maketitle

\begin{affiliations}
 \item Osservatorio Astrofisico di Torino, Istituto Nazionale di Astrofisica (INAF), I-10025 Pino Torinese, Italy
 \item Max Planck Institute for Astronomy, K\"onigstuhl 17, D-69117 Heidelberg, Germany
\end{affiliations}

\begin{abstract}

In a cosmological setting, the disc of a galaxy is expected to continuously experience gravitational torques and perturbations from a variety of sources, which can cause the disc to wobble, flare and warp\cite{Binney:1992,Sellwood:2013}. Specifically, the study of galactic warps and their dynamical nature can potentially reveal key information on the formation history of galaxies and the mass distribution of their halos. Our Milky Way presents a unique case study for galactic warps, thanks to detailed knowledge of its stellar distribution and kinematics. Using a simple model of how the warp's orientation is changing with time, we here measure the precession rate of the Milky Way's warp using 12 million giant stars from \gaia\ Data Release 2\cite{Brown:2018}, finding that it is precessing at $10.86 \pm 0.03_{stat} \pm 3.20_{syst}$ \kmskpc\ in the direction of Galactic rotation, about one third the angular rotation velocity at the Sun's position in the Galaxy. The direction and magnitude of the warp’s precession rate favour the scenario that the warp is the result of a recent or ongoing encounter with a satellite galaxy, rather than the relic of the ancient assembly history of the Galaxy.
\end{abstract}

\newpage
\section*{Main}

The Galactic warp is a large-scale distortion of the outer disc with respect to the inner disc, bearing witness that the disc of our Galaxy may be subject to external torques. Proposed mechanisms include torques from a misalignment of the disc's rotation axis with respect to the principle axis of a non-spherical halo\cite{Sparke:1988,Dubinski:2009}, or from accreted matter in the halo acquired during late infall\cite{Jiang:1999,Shen:2006}, or from nearby, interacting satellite galaxies and their consequent tides\cite{Weinberg:2006,Laporte:2019}. However, the cause and dynamical nature of the warp of our Galaxy has remained unclear due to a lack of kinematic constraints. Recent works using \gaia\ data have revealed the kinematic signature of the Galactic warp\cite{MWDR2:2018,Liu:2017,Schoenrich:2018,Poggio:2018,Romero:2019,Carrillo:2019,LopezCorredoira:2019}, while the geometry of the warp has been mapped on large-scales using Cepheids as tracers\cite{Chen:2019,Skowron:2019}. In this contribution we quantitatively analyse the warp kinematic signature using a simple kinematic model, with the purpose of constraining the possible mechanisms responsible for the warp of the Milky Way.

\begin{figure}
\begin{centering}
\includegraphics[width=1.05 \hsize]{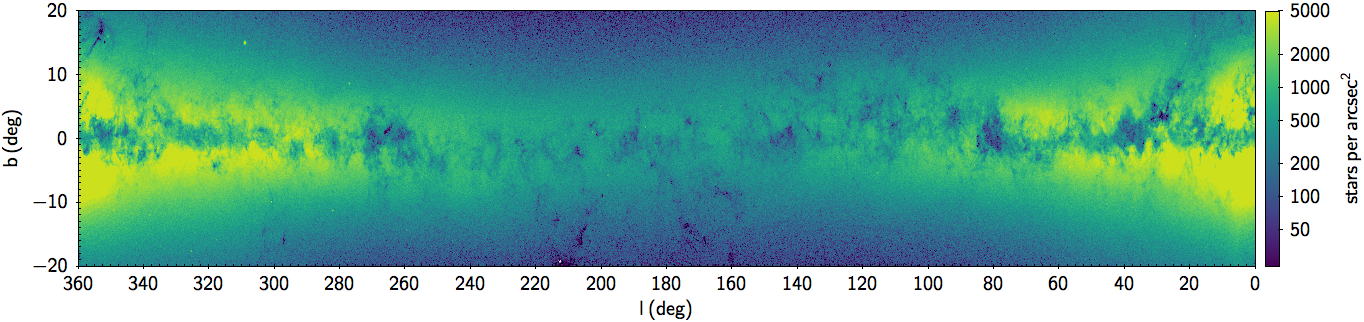} 
\includegraphics[width=1.05 \hsize]{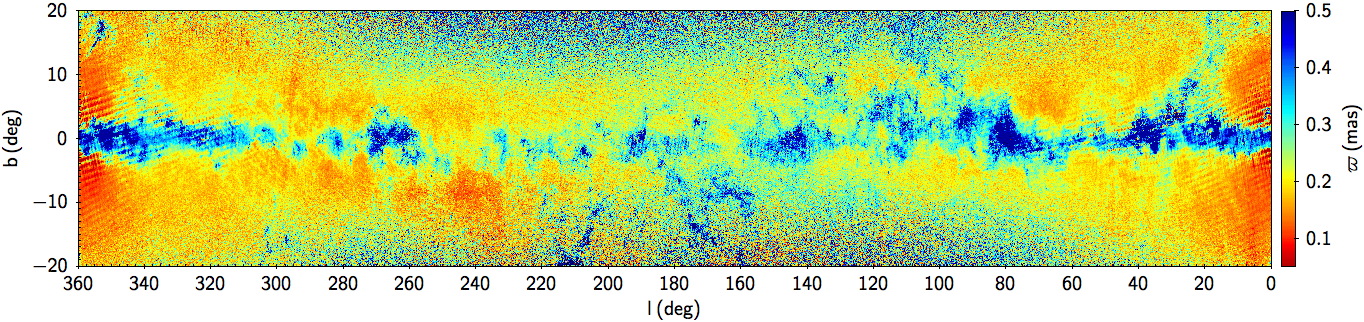}
\includegraphics[width=1.05 \hsize]{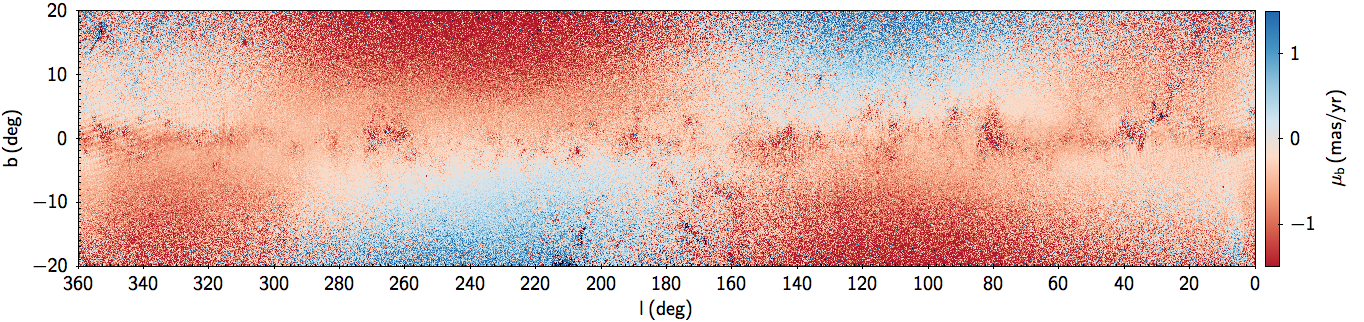}
\caption{{\bf Overview of the adopted dataset.} Sky-projection of the star counts (top), showing that the majority of the sample is concentrated in the Galactic plane. Extinction features caused by foreground dust clouds are clearly visible as under-dense regions. In these directions, where extinction limits the viewing distance, the median parallax $\varpi$ is typically larger (middle). The median proper motion $\mu_{b}$ on the sky (bottom) shows an alternate positive/negative pattern above and below the galactic plane, caused by the combination of differential Galactic rotation and solar motion. \label{Fig:data}}
\end{centering}
\end{figure}

The dataset for this study is the \emph{giant sample} from the second \gaia\ Data Release (\gdrtwo) presented in our previous work\cite{Poggio:2018}, selected via a probabilistic approach based on astrometry and photometry from \gdrtwo\, and 2MASS\cite{2MASS:2006}. The sample contains $12 \, 616 \, 068$ giant stars with apparent magnitude $G < 15.5$, and within $20^{\circ}$ from the galactic plane ($b=0^{\circ}$).  For each star we use \gdrtwo\ galactic coordinates $l$ and $b$, the parallax $\varpi$, and the component of the proper motion perpendicular to the galactic plane $\mu_{b}$, derived from \gaia's astrometry (see Methods). We only consider $\mu_{b}$ as the Galactic warp primarily manifests itself kinematically as systematic motions perpendicular to the Galactic plane\cite{Poggio:2017}. We also construct the covariance matrix $\Sigma_{\varpi,\mu_b}$ 
, which includes the astrometric uncertainties on $\varpi$ and $\mu_{b}$, together with their correlation $\rho_{\varpi,\mu_b}$, neglecting the uncertainties in the coordinates $(l,b)$. An overview of our dataset is provided in Figure \ref{Fig:data}. 

\begin{figure}
\begin{centering}
\includegraphics[trim={4cm 2cm 5cm 2cm},clip,width=1. \hsize]{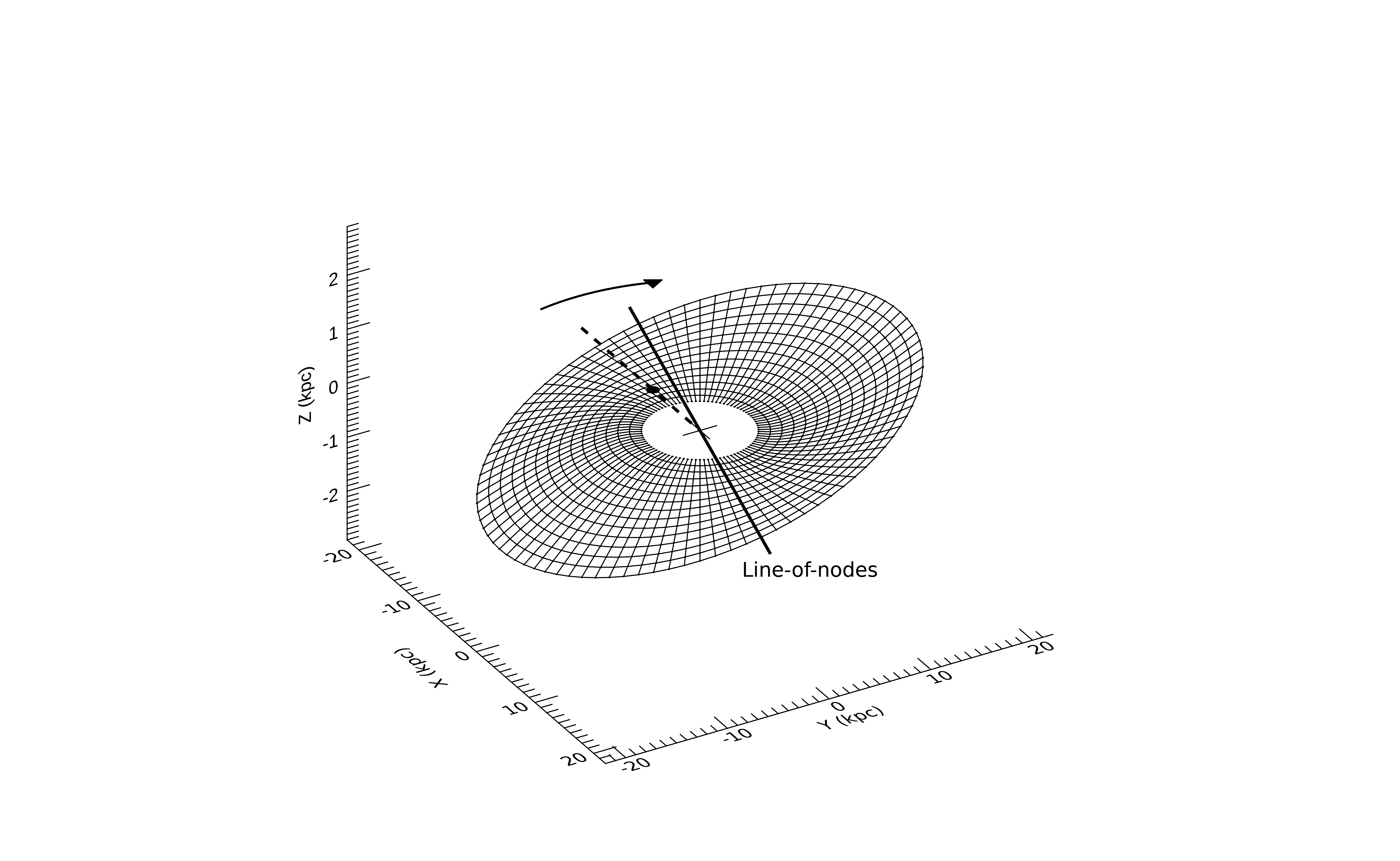} 
\caption{ {\bf Three-dimensional representation of the precessing warp model.} The Sun's position is shown by the dot, the Galactic center by the black cross, and the direction of Galactic rotation by the arrow. The angle between the warp's line-of-nodes (bold solid line) and the Sun's azimuth ($\phi=0^\circ$, dashed line) is the warp phase angle $\phi_{w}$ at the present time. 
  \label{Fig:model}}
\end{centering}
\end{figure}

The geometry of the warp can be simply described as a sinusoidal vertical displacement with respect to galactic azimuth, with an amplitude that gradually increases with galactocentric radius (see Figure \ref{Fig:model}). The line along which the vertical displacement is zero is referred to as the line-of-nodes. For a static warp geometry, the systematic vertical motions of the stars are expected only in the outer warped portions of the disc, in proportion to the warp amplitude, where stars must oscillate vertically as they rotate about the center of the Galaxy to maintain the geometry of the warp. 

As a first attempt to quantify the evolution of the warp we assume that the warp shape is constant, but consider the possibility that the orientation of the warp changes with time. In general there are two possible generic models. One is to consider that the warp rotates about an axis coincident with the vertical axis of the inner disc, whose orientation remains fixed. Such a model would describe the case where only the outer disc is somehow vertically perturbed (as might be expected from an interaction with a satellite galaxy), while the inner galaxy remains relatively unperturbed. A second possibility is that the entire inner disc, while remaining flat thanks to its own self-gravity, is precessing about some axis not coincident with the normal axis of the inner disc, due to a constant external torque, much like an inclined spinning top. This kinematic model would apply to those warp models invoking long-acting torques from a misaligned non-spherical halos or from a tilted outer torus of mass. In both cases the changing warp geometry can be simply described as a change in the direction of its line-of-nodes, parametrized by an angular rate $\omega$, traditionally referred to as the ``warp precession rate", a term inherited from tilted-ring models of galactic warps\cite{Sparke:1988}. For both of these kinematic models, the systematic vertical velocities resulting from Galactic rotation under the changing warp geometry are modulated with respect to those expected for a static warp, in proportion to $\omega$. However, as discussed further below, the expected warp precession rates of discs from models of non-spherical halos and tilted outer tori are very small, leading to imperceptibly small vertical motions in the inner disc, and thus consistent with a static warp model with $\omega=0$ given our uncertainties. For this reason we adopt a kinematic warp model of the first type, with a fixed orientation of the inner disc. See Methods for further details.

When the parameter $\omega$ is constant with respect to Galactocentric radius $R$ and $\omega>0$, the line-of-nodes remains straight and the warp rigidly ``precesses" in the same (prograde) direction as Galactic rotation. For the sake of simplicity we adopt a single value for $\omega$, in agreement with observations of external galaxies that show an initially straight line-of-nodes\cite{Briggs:1990}, and with theoretical studies that find the line-of-nodes is expected to be straight within  $R \lesssim 4.5$ disc scale lengths\cite{Shen:2006,BlandGerhard:2016}, which in our case encompasses approximately 90\% of our sample. In the presence of differential precession, our parameter $\omega$ can be interpreted as an average over the volume sampled by our dataset (see Methods).

An alternative model for a time dependent warp geometry might include a variable amplitude. We investigated the effect of a time-variable amplitude based on the observed warp geometries (see Methods), and found that such a model is not able
to reproduce the observed kinematic data (see Figure \ref{Fig:lmub}), unless one also imposes an amplitude much smaller than what is observed. We conclude that, if present, a time-variable warp amplitude is a second-order effect with respect to precession.

\begin{figure}
\begin{centering}
\includegraphics[width=0.7 \hsize]{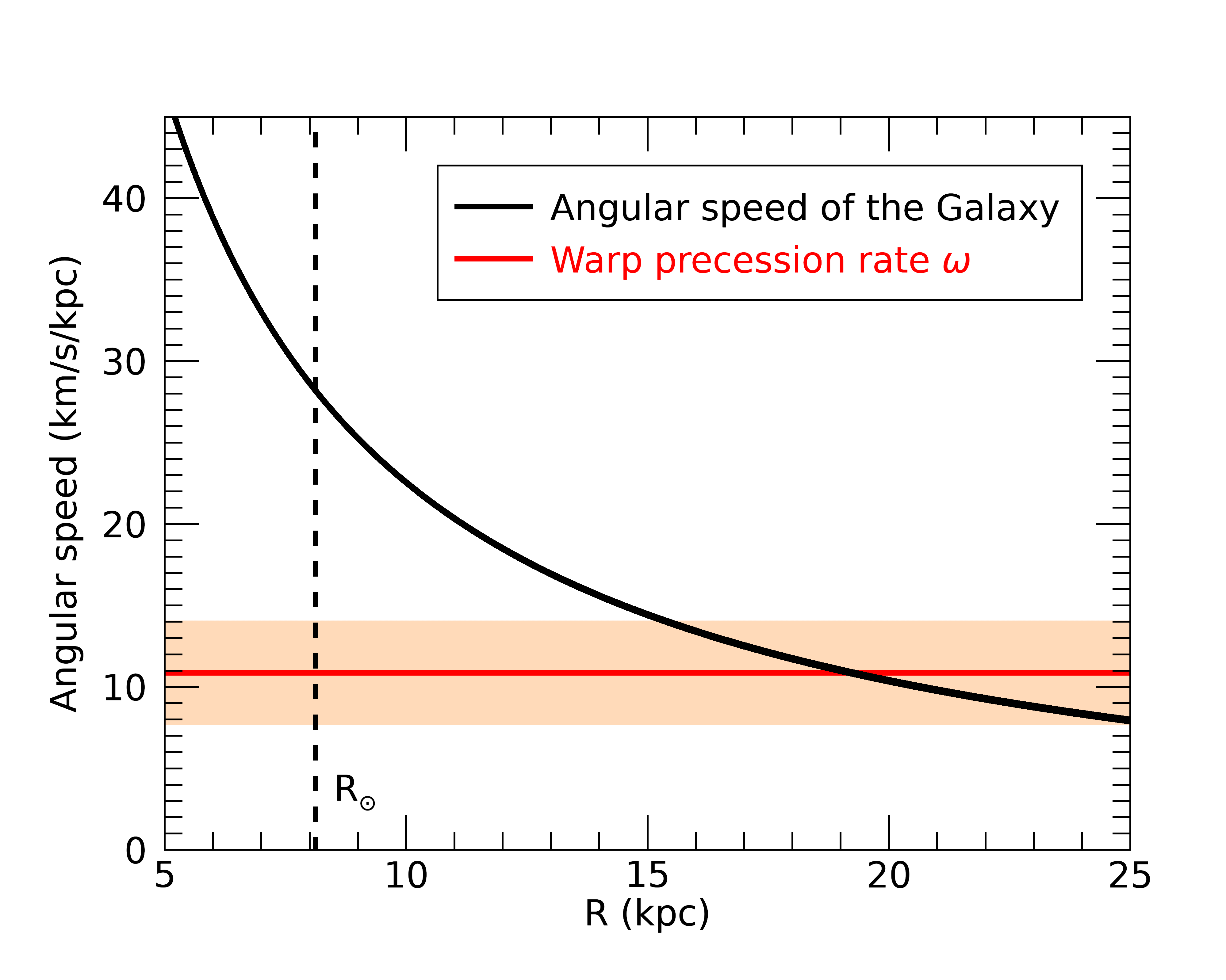} 
\caption{ {\bf Comparison of the warp precession rate $\omega$ and the angular speed of the Galaxy}, given by $\Omega(R)=V_C(R)/R$, where $V_C(R)$ is the circular velocity (see Methods). The dashed vertical line shows the position of the Sun ($R_{\odot}$), while the shaded area shows range of uncertainty on $\omega$ (random plus systematic). Extrapolating our obtained precession rate $\omega$ to the outer regions of the Galactic disc, the warp corotation radius (i.e. where the black and the red line cross) is at $\approx 20$ kpc with an uncertainty range from 15 to 25 kpc, though at these radii differential precession may be an issue. \label{Fig:corot}}
\end{centering}
\end{figure}

In this contribution, the warp precession rate $\omega$ is statistically inferred from the data assuming a model for the spatial and kinematic distribution of the giant stars in the Galaxy, taking into account the selection function of our sample, as well as a model for the measurement errors (the details are given in the Methods section). The inference is performed using the measured $(l,b,\varpi,\mu_b)$, and marginalizing over the unknown \emph{true} heliocentric distance $r$ and proper motion $\mu^\prime_{b}$. This procedure is done for four models of the warp geometry, as discussed in Methods. The average of the four obtained precession rates is 10.86 \kmskpc, indicating that the warp is precessing in the direction of Galactic rotation, at approximately one third of the angular velocity of the Galaxy at the location of the Sun. We find a statistical uncertainty of 0.03\kmskpc\ for all geometrical warp models. This uncertainty is quite small in part due to the large size of our dataset, and in part because in our evaluation of the posterior probability distribution function of $\omega$ all other parameters are fixed (assumed known) for each run. However, assuming different warp geometries that have been proposed for the stellar distribution leads to a variation in $\omega$ of several \kmskpc (see Extended Data Figure 3). This variation reflects in part our uncertain knowledge of the warp shape, and in part the fact that the different stellar populations considered here might have slightly different warp geometries\cite{Amores:2017}. The additional systematic uncertainties introduced by our uncertain knowledge of the Galaxy, including the warp geometry, are estimated by varying the most relevant parameters by their uncertainties and/or assuming different values available in the literature, and finding the resulting change in the estimated precession rate $\omega$ (see Methods for a detailed discussion). Our final estimate of the warp precession rate is:
\begin{equation}
\label{omega} 
\omega = 10.86 \pm 0.03_{stat} \pm 3.20_{syst} \text{\, km} \text{\, s}^{-1} \text{\, kpc}^{-1} \quad,
\end{equation}
where the value of $\omega$ is the average of the values quoted above for the four adopted warp geometries. For comparison, the pattern speed of the Galactic bar has been constrained to be $43 \pm 9$\kmskpc\cite{BlandGerhard:2016}, while that of the spiral pattern has been estimated to be between 17 and 28\kmskpc\ assuming the classical density wave theory\cite{Gerhard:2011}. Assuming rigid precession, the warp corotation radius is beyond 16 kpc (see Figure \ref{Fig:corot}), though at these radii differential precession may be an issue.

Compared with our dataset, we find that the precessing model gives a statistically better fit than a static model (see Methods), as is clearly shown in Figure \ref{Fig:lmub}. Alternatively, if we assume a static ($\omega=0$) warp and adjust its amplitude from the kinematics, we obtain a warp amplitude of 150 pc and 310 pc, respectively, at R=12 kpc and 16 kpc, about three times smaller than current estimates available in literature.

Our measurement of a significant prograde precession rate confirms that the Galactic warp is evolving with time, as already suggested by recent studies\cite{Amores:2017,Chen:2019}, and can be compared with the expectations of the warp mechanisms that have been proposed to date. Theoretical and numerical studies of misaligned non-spherical halo potentials find that the direction of the precession is determined by the shape of the halo potential, a prograde precession rate indicating that the halo is prolate or prolate-like rather than an oblate\cite{Sparke:1988,Ideta:2000,Jeon:2009,Dubinski:2009}. Recent works indicate that the Galactic stellar halo is dominated by the debris of an ancient major merging event with a massive dwarf galaxy (i.e. Gaia Enceladus\cite{Helmi:2018} or the ``Sausage" galaxy\cite{Belokurov:2018}), leading to a flattened triaxial (oblate-like) halo\cite{Iorio:2019}, which should produce a retrograde warp precession. However, the magnitude of our measured precession rate significantly exceeds all predictions from dynamical models involving torques from both prolate-like or oblate-like halo potentials, as well as those produced by a misaligned outer torus of later accreted material\cite{Jiang:1999,Shen:2006,Jeon:2009,Dubinski:2009}: typical precession periods from such models are approximately between 4 and 40 Gyr, corresponding to warp precession rates between 1.5 and 0.1 \kmskpc. Given our uncertainties, such small precession rates would be consistent with a static warp model. Whether such models can reproduce significantly larger precession rates under realistic assumptions should be explored in the future. 
 
The comparison with dynamical models available in the literature to date thus suggests that a large warp precession rate is more consistent with the warp being a transient response seen in the outer disc, rather than a slowly evolving structure from a misaligned non-spherical potential. Indeed, a recent interaction with a satellite galaxy could cause a more significant response of the outer Galactic disc\cite{Weinberg:2006,DOnghia:2016,Laporte:2019}. In this scenario our long-lived precessing warp model should not be taken as describing the global systematic vertical motions of the disc, but rather it provides a local approximation for the response of the outer Galactic disc to a recent perturbation, over the portion of the Milky Way disc covered by our dataset. Indeed, in the future more sophisticated time-dependent warp models should be explored, including differential precession and a time-dependent amplitude, both of which are expected features of transient warps. Nevertheless, our results suggests that external forces from interacting satellite galaxies are playing a significant and ongoing role in shaping the outer disc of the Milky Way.

\begin{figure}
\begin{centering}
\includegraphics[trim={0cm 0cm 0cm 0cm},clip,width=1  \hsize]{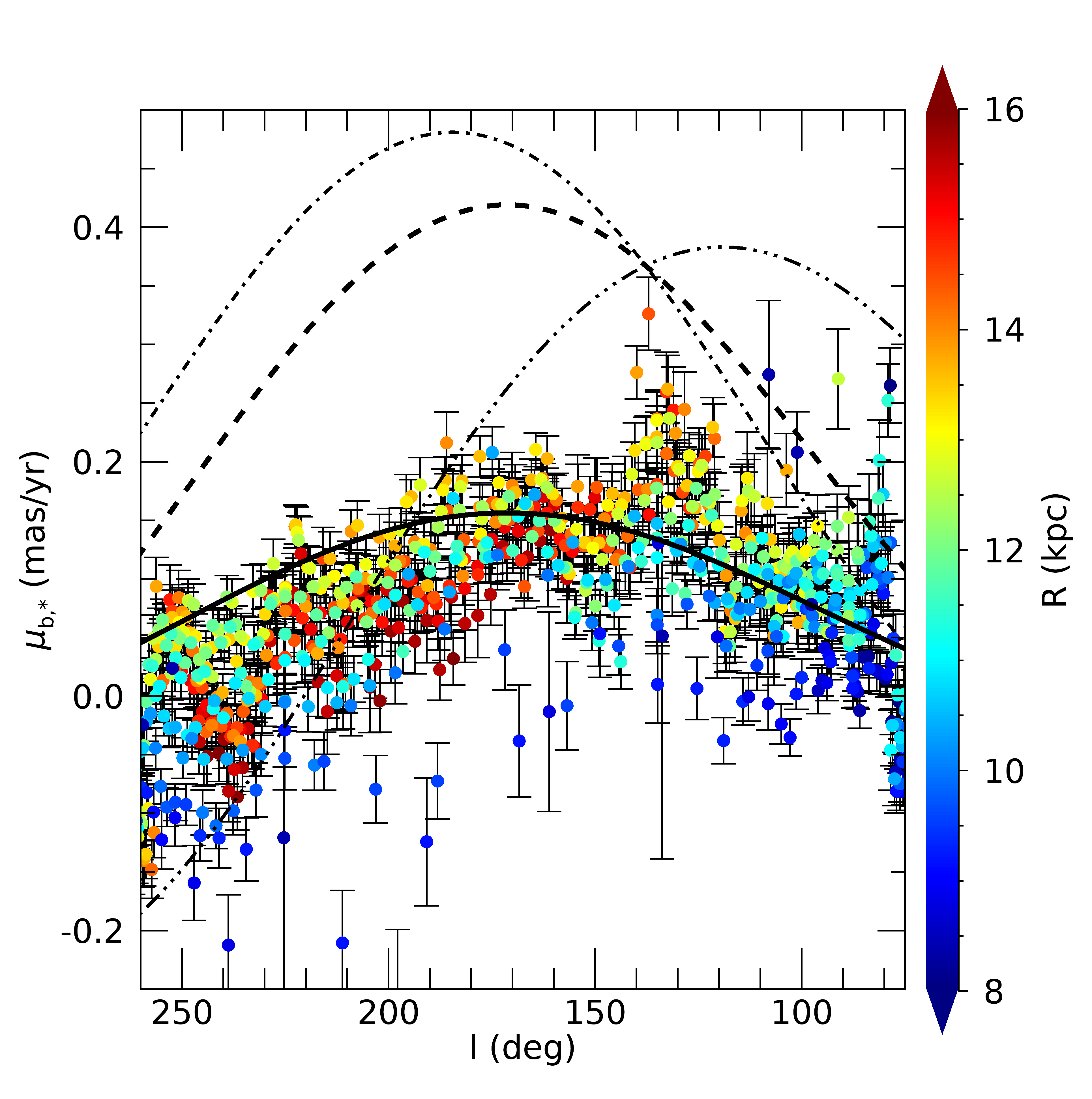} 
\caption{ {\bf Comparison of the data (colored points) and the models (lines).} Colored points show the median proper motions $\mu_{b,*}$
corrected for the Solar motion and Galactic rotation for 200 pc wide cells containing at least 500 stars, using distances
from our previous work\cite{Poggio:2018} (used here for illustrative purposes only). Error bars show 1-$\sigma$ bootstrap uncertainties on the median proper motion $\mu_{b,*}$. The dashed line corresponds to the
prediction from a static warp, while the solid line shows the
precessing model ( $\omega = 10$ \kmskpc),  assuming the geometry of
the Chen et al.\cite{Chen:2019} linear model (see Methods), both at R=13 kpc.  For comparison, we also show the prediction from a
time-varying amplitude model (see Equation \ref{vz_general} in
Methods) with $\partial h_w/\partial t$ = -5 (+15) kpc Gyr$^{-1}$ as a
dash dot (dash dot dot) line. Similar curves are found using the other
warp geometries considered. \label{Fig:lmub}}

\end{centering}
\end{figure}

\newpage

\begin{methods}

\renewcommand{\figurename}{ Extended Data Figure}

\subsection{The long-lived precessing warp model}

In general, the geometric shape of the Galactic warp can be modelled as a vertical displacement of the Galactic disc in Galactocentric cylindrical coordinates $(R,\phi,z)$, where $\phi$ is taken in the direction of Galactic rotation: 
\begin {equation}
\label{zwarp} 
z_w(R,\phi) = h_w(R) \, \sin{(\phi - \phi_{w}}) \quad,
\end{equation}
where $\phi_{w}$ is the phase angle of the warp defining the galactic azimuth of the warp's line-of-nodes, the line along which $z_w = 0$ (see Figure \ref{Fig:model}), and $h_w(R)$ is a height function specifying the maximum amplitude of the warp with respect to galactocentric radius $R$ (see Equation \ref{hzwarp} below). We have adopted here four models of the warp geometry: two recently presented by Chen et al.\cite{Chen:2019} based on a sample of Cepheids, one by Yusifov\cite{Yusifov:2004} based on pulsars, and another one by L\'{o}pez-Corredoira et al.\cite{LopezCorredoira:2002B} based on 2MASS red-clump stars. In Momany et al.\cite{Momany:2006}, the Yusifov model has been found to describe well the RGB stars, a population similar to the sample used in this work. The warp geometry parameters are listed in Extended Data Figure 3. 

In order to describe the temporal evolution of the warp, we can model a change in Galactic azimuth of the line-of-nodes $\phi_w(t) = \phi_{0,w}  + \omega \, t$, where $\phi_w = \phi_{0,w}$ is the current azimuth of the line-of-nodes, and the variable $\omega$ is the warp precession rate. To have a general approach (that will be simplified later), $\omega$ can be a function of R, and the warp amplitude $h_w(R,t)$ can vary with time\cite{LopezCorredoira:2014}. Following Binney \& Tremaine\cite{Binney:2008}, we can model our Galaxy as a collisionless system, and take the $0^{th}$ moment of the Collisionless Boltzmann Equation
\begin {equation}
\label{zeromom}
\frac{\partial \rho}{\partial t} + \sum_{i=1}^{3} \frac{\partial (\rho \overline{V}_i )}{\partial x_i} = 0 \quad ,
\end{equation}
which in Galactocentric cylindrical coordinates $(R, \phi, z)$ becomes (assuming that we have no mean radial motion, see below): 
\begin {equation}
\label{zeromom_cyl}
\frac{\partial \rho}{\partial t}+ \frac{1}{R} \frac{\partial (\rho \overline{V}_\phi)}{\partial \phi}
                                               + \frac{\partial (\rho \overline{V}_z)}{\partial z} = 0 \quad .                        
\end{equation}                                               
This equation can be further simplified if we assume that the azimuthal velocity only depends on $R$ and that the mean vertical velocity does not depend on z:
\begin {equation}
\label{zeromom_fin}
\frac{\partial \rho}{\partial t}  + \frac{\overline{V}_\phi}{R} \frac{\partial \rho }{\partial \phi}
                                               + \overline{V}_z\frac{\partial \rho }{\partial z} = 0 \quad .                                                                  
\end{equation}
We can therefore calculate the terms $\frac{\partial \rho}{\partial t} $, $\frac{\partial \rho }{\partial \phi} $ and $\frac{\partial \rho }{\partial z} $ from the assumed density of the Galactic disc (Equation \ref{density}), which is warped according to Equation \ref{zwarp}, so that Equation \ref{zeromom_fin} can be solved with respect to $\overline{V}_z$, obtaining 
\begin {equation} 
\label{vz_general} 
\overline{V}_{z} (R,\phi, t=0)=\left(\frac{\overline{V}_\phi}{R} - \omega(R) \right) \, h_w(R) \, \cos{( \phi - \phi_w)} + \frac{ \partial h_w}{\partial t} \sin{( \phi - \phi_w)}
\end{equation}
where $\overline{V}_\phi $ is the mean azimuthal velocity. As discussed in the main text and shown in Figure \ref{Fig:lmub}, setting $\omega = 0$ and varying $\frac{ \partial h_w}{\partial t}$ cannot reproduce the observed kinematic signature of the warp. For this reason, we conclude that the most relevant term is the one containing the precession $\omega$, and neglect the term $\frac{ \partial h_w}{\partial t}$. Moreover, as discussed above, the simplest model for warp precession is a rigidly precessing warp with $\omega$ independent of $R$, obtaining
\begin {equation} 
\label{vz_precessing} 
\overline{V}_{z} (R,\phi)=\left(\frac{\overline{V}_\phi}{R} - \omega \right) \, h_w(R) \, \cos{( \phi - \phi_w)} \quad.
\end{equation}
As a test, we estimated $\omega$ for subsets of our data in overlapping rings of a few kiloparsecs in $R$, and found the variation of $\omega$ with respect to $R$ to be of the same order of our systematic uncertainties (Equation \ref{omega}); for this reason, we here assume a single value for $\omega$, and leave the exploration of possible trends of $\omega$ as a function of $R$ for future works. Nevertheless, we found that the measured precession rate in each radial bin is significantly different from zero and consistent with the estimate calculated from the whole sample. As expected, averaging over the different rings is equivalent to assume a single value for $\omega$. Finally, it has been shown that a lopsided warp can also cause a decrease in the observed kinematic signal\cite{Romero:2019}. However, the warp geometries adopted here describe the warp to be symmetric on both sides of the Galaxy; as noted by Momany et al.\cite{Momany:2006}, such models can reproduce an asymmetry in the observed stellar distribution if the Sun is not lying on the line-of-nodes\cite{Yusifov:2004,Chen:2019,Romero:2019}.

\subsection{The statistical inference}

To infer the warp precession rate $\omega$, we calculate the likelihood of our dataset given the model and the uncertainties by marginalizing over the \emph{true} unkown quantities, namely the \emph{true} heliocentric distance $r$ and the \emph{true} proper motion $\mu^\prime_{b}$. In the following, we use bold font for vectors and matrices. As discussed below, the model parameters (collected into the symbol $\boldsymbol{\Theta}_{Gal}$) are maintained fixed, while $\omega$ is the free parameter that must be estimated. Assuming that the galactic coordinates $(l,b)$ are noise-free quantities, the likelihood of the $i^{th}$ star of our sample can be written for a given value of the warp precession rate $\omega$:
\begin{equation}
\label{big_int}
      p_i( \varpi, \mu_{b}  \, |\,   l, b, \boldsymbol{   \Sigma_{\varpi,\mu_b}}  ,\boldsymbol{\Theta}_{Gal}, \omega )   
       = 
           \iint 
       p_i(\varpi, \mu_{b}  |\, \boldsymbol{   \Sigma_{\varpi,\mu_b}},  r, \mu^\prime_{b} \,) \, 
       p_i( r, \mu^\prime_{b} \, |\, l, b, \boldsymbol{\Theta}_{Gal}, \omega) \, d{\mu^\prime_{b}} \, d{r}
          \quad ,     
\end{equation}
which represents the probability of a star at $(l, b)$ to have a measured $\varpi$ and $\mu_{b}$ according to the measurement model $p_i(\varpi, \mu_{b}  |\, \boldsymbol{   \Sigma_{\varpi,\mu_b}},  r , \mu^\prime_{b} \,) $ and the model of the Galaxy $ p_i( r, \mu^\prime_{b} \, |\, l, b, \boldsymbol{\Theta}_{Gal}, \omega)$, which includes the \emph{noise-free} spatial and kinematic distribution of the giant stars in the Galaxy, as well as the modeled selection function of our sample. (The details are given in the following sections.) Once the likelihood is obtained for each star, we can calculate the likelihood of the dataset as
\begin {equation}
\label{product} 
p( \{ \varpi, \mu_{b} \}_N \, |\{ \, l, b, \, \boldsymbol{   \Sigma_{\varpi,\mu_b}} \}_N ,\boldsymbol{\Theta}_{Gal}, \omega )  = 
\Pi_i^N p_i( \varpi, \mu_{b}  \, |\, l, b,  \boldsymbol{   \Sigma_{\varpi,\mu_b}}  ,\boldsymbol{\Theta}_{Gal}, \omega ) \quad,
\end{equation}
for a given $\omega$, where the product is over the $N$ stars of our sample, assuming that each star's data is an independent draw from a parent distribution given by our model. Repeating the above procedure over a range of values of $\omega$, we find the likelihood of our dataset as a function of $\omega$, which is proportional to the posterior probability distribution function (pdf) of $\omega$ given our dataset if a uniform prior is assumed.

\subsection{The model of the Galaxy}

Our assumed model of the Galaxy is specifically designed to include the properties of the adopted dataset, which is expected to contain mostly thin disc giant stars\cite{Poggio:2018}. Treating our sample as a simple stellar population, we can write 
\begin{equation}
\label{model_int}
        p_i( r, \mu^\prime_{b} \, |\, l, b, \boldsymbol{\Theta}_{Gal}, \omega)  = 
        p_i( \mu^\prime_b \, | \, l, b, r, \boldsymbol{\Theta}_{KIN}, \omega)  \, p_i( r \, | \, l, b,\boldsymbol{\Theta}_{SP},\boldsymbol{\Theta}_{CMD})
       \quad,
\end{equation}
where $p_i( \mu^\prime_b \, | \, l, b, r, \boldsymbol{\Theta}_{KIN}, \omega)$ is the probability density function for the \emph{true} proper motion $\mu^\prime_{b}$ for a given \emph{true} position $(l,b,r)$ according to the kinematic model $\boldsymbol{\Theta}_{KIN}$, and $p_i( r \, | \, l, b,\boldsymbol{\Theta}_{SP},\boldsymbol{\Theta}_{CMD})$ is the expected probability density function of the heliocentric distance $r$ in the direction $(l,b)$ according to the assumed spatial model $\boldsymbol{\Theta}_{SP}$, taking into account the spatial observational effects due to the selection function of our sample, collected in the symbol $\boldsymbol{\Theta}_{CMD}$ (all these terms are described in the following). Our separation of the Galaxy model into a kinematic and spatial component in Equation \ref{model_int} permits us to solve analytically the integral over $\mu_b$ in Equation \ref{big_int} for a given position $(l,b,r)$ in the Galaxy, while the one over heliocentric distance $r$ is calculated numerically via Monte Carlo integration. For a star at $(l,b)$, the probability density function (pdf) of $r$ along the line-of-sight can be written as
\begin{equation}
\label{model_int_selfun}
        p( r \, | \, l, b,\boldsymbol{\Theta}_{SP},\boldsymbol{\Theta}_{CMD})  = 
                    p( r \, | \, l, b,\boldsymbol{\Theta}_{SP}) S( r \, | \, \boldsymbol{\Theta}_{CMD})
\end{equation}
which is composed of the pdf of $r$, $p( r \, | \, l, b,\boldsymbol{\Theta}_{SP})$, given the assumed model of the spatial distribution of stars in the Galaxy (see next section), and the term $S( r \, | \,\boldsymbol{\Theta}_{CMD})$, namely the selection function specifying the fraction of stars that can be observed at $r$, according to the modelled distribution of stars in color-magnitude space, our selection criteria and the completeness of our sample, captured collectively by the symbol $\boldsymbol{\Theta}_{CMD}$.  We clarify that $S( r \, | \, \boldsymbol{\Theta}_{CMD})$ is not a pdf, as discussed in the following.

\subsection{The selection function \label{selfun}}

For a magnitude-limited sample, the fraction of observable objects decreases as a function of heliocentric distance $r$. To reproduce this observational effect, we need a model for the luminosity function $\Phi(M_G)$, where $M_G$ is the absolute magnitude in the $G$ band. In order to take into account extinction, we instead consider the distribution of stars with respect to the absolute magnitude uncorrected for extinction in the $G$ band, $f_{A_V}(\mathcal{M}_G)$, where $\mathcal{M}_G \equiv (M_G + A_G)$, $A_G$ being the extinction in the $G$ band.  The subscript $A_V$ reminds us that this distribution depends on the amount of extinction, which we parameterize with $A_V$, since our adopted extinction map\cite{Drimmel:2003} specifies the extinction in the $V$ band. We derive $f_{A_V}(\mathcal{M}_G)$ by first constructing the distribution of stars in color-magnitude space (i.e. a Hess diagram) from the PARSEC isochrones\cite{Bressan:2012, Chen:2014,Chen:2015, Tang:2014}, assuming a constant star formation rate, the canonical two-part power law Kroupa initial mass funtion corrected for unresolved binaries\cite{Kroupa:2001, Kroupa:2002}, and solar metallicity, for a given value of $A_V$. Finally, the distribution of stars in observed colour space is modified by applying the selection criteria of our sample (see Poggio et al\cite{Poggio:2018}), i.e. by setting to zero the parts of the distribution not satisfying the adopted colour-colour cuts. Using this procedure we construct a suite of  ``luminosity" functions $f_{A_V}(\mathcal{M}_G)$ for different values of $A_V$. For computational convenience, from each $f_{A_V}(\mathcal{M}_G)$ we derive the cumulative distribution function: 
\begin{equation}
F_{A_V} (\mathcal{M}_G) = \frac{\int_{-\infty}^{\mathcal{M}_G} f_{A_V} (\mathcal{M}^\prime_G) d\mathcal{M}^\prime_G}
           {\int_{-\infty}^{+\infty} f_{A_V} (\mathcal{M}^\prime_G) d\mathcal{M}^\prime_G} \quad, 
\end{equation}
where the prime indicates the variable of integration. (See Extended Data Figure 1.)

Finally, we include in our model the completeness of our dataset, which was extracted from a combined Gaia DR2$\cap$2MASS catalogue\cite{Poggio:2018}. The completeness of this catalogue is evaluated by assuming that GaiaDR2 and 2MASS are independent samples of the sky, that is, that the completeness $C_j(G_j)$ of the DR2$\cap$2MASS for the $j^{th}$ magnitude bin $G_j$ is approximately equal to $N_{GT}^2/N_G N_T$, where $N_G$ and $N_T$ are the counts of DR2 and 2MASS sources in the $G_j$ bin, and $N_{GT}$ the number of sources in both DR2 and 2MASS in the same $G_j$ bin. (See Drimmel et al for details.) By definition, the completeness $C_j(G_j)$ lies between 0 and 1 (see Extended Data Figure 2). 

Thus, for a given position $(l,b,r)$, we recover the extinction in the V band, $A_V$, from the adopted extinction map\cite{Drimmel:2003}, which in turn allows us to choose an appropriate $f_{A_V}(\mathcal{M}_G)$ that takes into account our selection criteria. Finally, the fraction of stars at $r$ that are observable within our apparent magnitude limits $(5 < G < 15.5)$ is found by integrating $f_{A_V}(\mathcal{M}_G)$ modulated by the completeness of our sample over the appropriate range of $\mathcal{M}_G(G,r) =  G - 5 \log_{10}{r} + 5 $, similar to Astraatmadja \& Bailer-Jones\cite{Astr:2016} and Bailer-Jones\cite{BailerJones:2018c}:
\begin{equation}
S(r| \, \boldsymbol{\Theta}_{CMD}) = \int_{G=5}^{15.5} C(G) f_{A_V}(\mathcal{M}_G(G,r)) dG \approx \sum_j C_j I_j
\end{equation}
where the sum is over apparent magnitude bins of width $\Delta G = 0.5$ magnitude in $G$, and
\begin{equation}
I_j = \int_{\mathcal{M}^-_G}^{\mathcal{M}^+_G} f_{A_V}(\mathcal{M}_G) d\mathcal{M}_G = F_{A_V}(\mathcal{M}^+_G) - F_{A_V}(\mathcal{M}^-_G)\, ,
\end{equation}
where $\mathcal{M}^\pm_G = G_j \pm \Delta G/2 - 5 \log_{10}{r} + 5$ are the upper/lower limits of the $j^{th}$ magnitude bin.  

\subsection{The spatial model \label{sp}} 
For a given direction $(l,b)$, the probability of a star to be at a \emph{true} distance $r$ is proportional to the stellar density $\rho$ predicted by our spatial model $\boldsymbol{\Theta}_{SP}$, so that we can write $ p_i(r \, | \, l, b, \boldsymbol{\Theta}_{SP}) \propto r^2 \,\rho(r \, | \, l, b, \boldsymbol{\Theta}_{SP} ) $, where the term $ r^2$ takes volume effects into account. We now move from heliocentric $(l, b, r)$ to Galactocentric coordinates $(R,\phi, z)$, which are convenient for describing our Galactic model. The stellar density is modelled as an exponential disc
\begin {equation}
\label{density} 
\rho(R,\phi, z) = \rho_0  e^{-(R - R_{\odot})/L_R}  e^{-(|z- z_w|/h_z(R))}\quad,
\end{equation}
with scale length $L_R$ and scale height
\begin {equation}
\label{flare} 
h_z(R) = h_z(R_{\odot}) e^{(R - R_{\odot})/h_{fl}}\quad,
\end{equation}
which increases as a function of R because of the disc flaring.  Meanwhile the Galactic disc is warped according to Equation \ref{zwarp},
where the height function $h_w(R)$ is the given by:
\begin {equation}
\label{hzwarp} 
h_w(R)=h_{0,w} \, (R-R_w)^{\alpha_w} 
\end{equation}
for $R > R_w$ and zero otherwise, and the parameters $h_{0,w}$ and $\alpha_w$ modulating the magnitude and the rate at which the warp amplitude increases with galactocentric radius.
Several parametrizations are available in literature\cite{Skowron:2019,Chen:2019,Reyle:2009,Robin:2008,Marshall:2006,Yusifov:2004,Drimmel:2001}. We here focus on the spatial warp models collected in Extended Data Figure 3. The remaining spatial parameters of the Galaxy model are reported in Extended Data Figure 4.


\subsection{The kinematic model \label{kin}}

The kinematic term $ p_i(\mu^\prime_b \, | \, l, b, r, \boldsymbol{\Theta}_{KIN}, \omega) $ in Equation \ref{model_int} predicts the probability of a star at $(l, b, r)$ to have a \emph{true} proper motion $\mu^\prime_b$ according to our kinematic model $\boldsymbol{\Theta}_{KIN}$ and warp precession $\omega$. Extended Data Figure 4 shows the kinematic parameters adopted in this work. Approximating the local velocity distribution as a velocity ellipsoid, one can show that
\begin {equation}
\label{mub_prob} 
p_i(\mu^\prime_b \, | \, l, b, r, \boldsymbol{\Theta}_{KIN}, \omega) = \frac{1}{\sqrt{2 \pi} \sigma^{\prime}_{\mu_b}} \, e^{- \frac{1}{2} \Bigl( \frac{\mu^\prime_b - \bar{\mu}^{\prime}_{b} }{\sigma^{\prime}_{\mu_b}}  \Bigr)^2 } \quad,
\end{equation}
where the mean value $\bar{\mu}^{\prime}_{b}$ is determined by the mean stellar velocities $(\overline{V}_R, \overline{V}_{\phi}, \overline{V}_{Z})$ in a given position $(l,b,r)$, together with the solar velocity with respect to the Galactic Center
\begin {equation}
\label{mub_th} 
\bar{\mu}^{\prime}_{b} = \frac{0.2104}{ r}  \Bigl[  \Bigl(\overline{V}_Z - V_{Z,\odot} \Bigr) \cos{b} +  \Bigl(\overline{V}_R - V_{R,\odot} \Bigr) \sin{b} \cos{(l + \phi)} -  \Bigl(\overline{V}_{\phi} - V_{\phi,\odot} \Bigr) \sin{b} \sin{(l + \phi)} \Bigr] \quad,
\end{equation}
while the intrinsic dispersion in the proper motions $\sigma^{\prime}_{\mu_b}$ is given by the intrinsic velocity dispersions and their correlations
\begin {equation}
\label{mub_sigma} 
\sigma^{\prime}_{\mu_b} = \frac{0.2104}{ r} \sqrt{ \sigma^2_{V_Z}\cos^2{b} + \sin^2{b} \Bigl[ \sigma^2_{V_R} \cos^2{(l+\phi)} + \sin^2{(l+\phi)} ( \sigma^2_{V_{\phi}}  
         + 2 \overline{V_R V_Z} \cos{(l+\phi)}) } \Bigr]
\end{equation}
which are described in the following.
The velocity dispersions exponentially decrease as a function of $R$ with a scale length $L_{\sigma}$, e.g. $\sigma_{R}(R)=\sigma_{R,0}\exp{(-(R - R_{\odot})/L_\sigma)}$, whereas the covariance $\overline{V_R V_Z}$ is characterised by the tilt angle $\alpha_{tilt}$ of the velocity ellipsoid 
\begin {equation}
\label{alpha_tilt} 
\overline{V_R V_Z} =\frac{(\sigma^2_R - \sigma^2_Z)}{2} \tan(2 \alpha_{tilt})\quad,
\end{equation}
where we adopt $\alpha_{tilt} = -0.9 \arctan(|z|/ R_{\odot}) - 0.01$\cite{Buedenbender:2015}, which is close to the case of alignment with the Galactocentric spherical coordinate
system. The mean azimuthal velocity is taken as $\overline{V}_{\phi}(R)=V_C(R) - V_A(R)$, where $V_C (R) = 229 - 1.7\, (R - R_{\odot})$ km/s\cite{Eilers:2019} is the mean circular velocity, and $V_A(R)$ is the asymmetric drift\cite{Binney:2008}
\begin {equation}
\label{asymm_drift} 
V_A(R)=  \frac{\sigma^2_R(R)}{2 \, V_C(R)} \Bigl[  \frac{\sigma^2_{\phi}(R)}{\sigma^2_R(R)} - 1 + R \Bigl( \frac{1}{L_R} + \frac{2}{L_{\sigma}} \Bigr) - \frac{R}{\sigma^2_R(R)} \frac {\partial \overline{V_R V_Z}}{\partial z}\Bigr] \quad.
\end{equation}
According to our simple model, there is no net radial motion, i.e. $\overline{V}_R=0$ and no vertex deviation. Finally, the warping of the disc along the $z$-coordinate (see Equations \ref{zwarp} and \ref{hzwarp}) induces the systematic vertical velocity $\overline{V}_Z$ described by equation \ref{vz_precessing}.
Thus, the warp manifests itself as a modulation of the mean vertical velocities with respect to $\phi$ and $R$. As expected, setting $\omega=0$ in Equation \ref{vz_precessing} recovers the vertical velocities for a static warp model\cite{Poggio:2017}. 



\subsection{The measurement model \label{noise}}
The term  $  p_i(\varpi, \mu_{b}  |\, \boldsymbol{   \Sigma_{\varpi,\mu_b}},  r, \mu^\prime_{b} \,) $ in Equation \ref{big_int} models the noise of the measurement process for the astrometric observables $(\varpi, \mu_{b})$ of the $i^{th}$ star of the catalogue. As reported in the text, we assume that the uncertainties on the position $(l,b)$ are insignificant compared to those of the astrometric quantities $\varpi$ and $\mu_b$, so that the non-trivial elements of the astrometric covariance matrix $\boldsymbol{\Sigma_{\varpi,\mu_b} }$ are 
\begin{equation}
\label{coV_matrix}
\boldsymbol{\Sigma_{\varpi,\mu_b} } =
\begin{pmatrix}
  \sigma^2_{\varpi}      & \rho_{\varpi \mu_{b}} \sigma_{\varpi}  \sigma_{\mu_{b}}  \\
  \rho_{\varpi \mu_{b}} \sigma_{\varpi}  \sigma_{\mu_{b}}                 & \sigma^2_{\mu_{b}}  
\end{pmatrix} \quad,
\end{equation}
which is constructed using the uncertainty on the measured parallax $\sigma_{\varpi}$ and proper motion $\sigma_{\mu_{b}}$, together with their correlation $\rho_{\varpi \mu_{b}}$. Since the covariance matrix was provided for each \gdrtwo\, source in equatorial (ICRS) coordinates, we converted it into galactic coordinates using the rotation matrix from the IAU SOFA Software Collection\cite{SOFA:2018-01-30}, according to the procedure described in Section 1.5 of ``The Hipparcos and Tycho Catalogues"\cite{ESA:Hipparcos}. The obtained covariance matrix uniquely defines the bivariate gaussian
\begin{equation}
\begin{split}
\label{mubplxcorr_2dgauss}
        p_i(\varpi, \mu_{b}  |\, \boldsymbol{   \Sigma_{\varpi,\mu_b}},  r, \mu^\prime_{b} \,)        & = \frac{1}{2 \pi \sigma_{\varpi} \sigma_{\mu_b} \sqrt{1 - \rho^2_{\varpi \mu_{b}}}} e^{-\frac{1}{2} 
          \Bigl[ 
            \frac{   \Delta_{\mu_b} ^2}{ (1-\rho^2_{\varpi \mu_{b}})  \sigma^2_{\mu_b}}  
           - \frac{ 2 \rho_{\varpi \mu_{b}}\, \Delta_{\mu_b} \Delta_{\varpi}   }{ (1-\rho^2_{\varpi \mu_{b}}) \sigma_{\mu_b} \sigma_{\varpi}  }  
           +  \frac{ \Delta_{\varpi} ^2 }{  (1-\rho^2_{\varpi \mu_{b}})  \sigma^2_{\varpi}} 
            \Bigr]
          } 
          \quad ,
\end{split}          
\end{equation}
with $\Delta_{\mu_b} = \mu^\prime_b - \mu_{b}$ and $\Delta_{\varpi} = 1/r + \varpi_{0} - \varpi $, where $\varpi_{0}$ is the systematic zeropoint offset of the \gdrtwo\, parallaxes. Unfortunately, no unique $\varpi_{0}$ is universally valid for all \gdrtwo\, stars, as it likely depends on magnitude, position and, possibly, color. Numerous estimates are available in literature\cite{Lindegren:2018,Arenou:2018,Riess:2018,Zinn:2019,Schoenrich:2019}, most of them lying between -0.03 and -0.08 mas. In this work, we adopt a value of $\varpi_{0}=-0.05$ mas, and discuss the impact of the assumed $\varpi_{0}$ on our results (see below). To take into account the additional uncertainties introduced by the variation of the systematic errors on small angular scales, we inflate the parallax and proper motion uncertainties $\sigma_{\varpi},\sigma_{\mu_{b}}$ with respect to the formal uncertainties $\tilde{\sigma}_{\varpi},\tilde{\sigma}_{\mu_{b}}$ obtained from the \gdrtwo\, catalogue 
\begin{equation}
\label{syst}
\sigma_{\varpi} =\sqrt{ k^2 \tilde{\sigma}^2_{\varpi} + \sigma^2_{\varpi,syst}} \quad,
\end{equation}
and similarly for $\sigma_{\mu_b}$, assuming $k=1.1$, $\sigma_{\varpi,syst}=0.043$ mas and $\sigma_{\mu_b,syst}=0.066$ mas/yr, as suggested by Lindegren ({\small \url{https://www.cosmos.esa.int/documents/29201/1770596/Lindegren_GaiaDR2_Astrometry_extended.pdf/1ebddb25-f010-6437-cb14-0e360e2d9f09})} for the stars with $G>13$, which applies to most of the stars in our dataset. 


\subsection{Systematic uncertainties on the warp precession rate}
We here discuss the impact of the systematic uncertainties introduced by our uncertain knowledge of the Galaxy. As seen in Equation \ref{vz_precessing}, there is an intimate connection between the adopted geometrical model of the warp and the corresponding kinematic signature. In particular, the warp precession $\omega$ and the spatial amplitude of the warp have similar effects on the kinematics, as they both modulate the amplitude of the signal in the mean vertical velocities. Between the different warp geometries considered, we find the range of derived warp precession rates to be 2.19\kmskpc. Other parameters of our model $\boldsymbol{\Theta}_{Gal}$ also influence our measurement of the warp precession rate, and their impact is estimated by varying each parameter by their uncertainties and/or assuming different values available in the literature, and repeating the procedure to estimate $\omega$ (see Main), again for each of the assumed warp geometries. We find that the most relevant contributions to the uncertainty of $\omega$ are the velocity dispersion at the Sun\cite{BlandGerhard:2016} $(\sigma_{(R,0)},\sigma_{(Z,0)})$, causing a variation in $\omega$ of 0.81\kmskpc, and the circular velocity at the Solar circle $V_C (R_{\odot})$, whose variation from 217\cite{Wegg:2019} to 240\cite{Reid:2014} km/s results in a variation of 1.55\kmskpc in $\omega$. Moreover, by varying the modelled systematic parallax zero-point error $\varpi_{0}$ (see above) between $-0.03$ and $-0.08$ mas we find a difference of 1.54 \kmskpc\ in the estimated $\omega$. By summing in quadrature the uncertainties just mentioned above, we obtain the systematic uncertainty 3.20 \kmskpc, which is reported in Equation \ref{omega}. As a final test, we also performed the analysis with the faint ($G>13$) sources only, in order to check that the spin of the bright reference frame of \gdrtwo\ \cite{Lindegren:2020} does not significantly affect our final estimate of warp precession rate.


\subsection{Statistical comparison of the static and precessing model}

We here compare the two models $M_{static}$ and $M_{precessing}$ and calculate which one is in better agreement with our dataset D. Following Equation 11.16 of `` Practical Bayesian Inference'' \cite{BailerJones:2017}, we calculate the probability $p(D|M_{precessing})$ using a Monte Carlo approach, drawing samples of $\omega$ from a broad uniform prior (here taken to be between -100 and 100 \kmskpc). Similarly, we calculate $p(D|M_{static})$, but now within a range of $\pm 3.20$ (which corresponds to our uncertainty) from $\omega=0 $ \kmskpc. The Bayes Factor for the two models is
\begin {equation}
\label{BF}
BF=\frac{p(D|M_{static})}{p(D|M_{precessing})} = \exp(-1389) \quad,
\end{equation}
which is significantly favouring the precessing model. Similar results are obtained also increasing or decreasing the prior range for $p(D|M_{static})$ by a factor of two.


\newpage

\section*{Extended data}

\begin{figure}[hb!]
\setcounter{figure}{0}
\begin{centering} 
\includegraphics[trim={0.4cm 0.5cm 1cm 1cm},clip,width=0.42  \hsize]{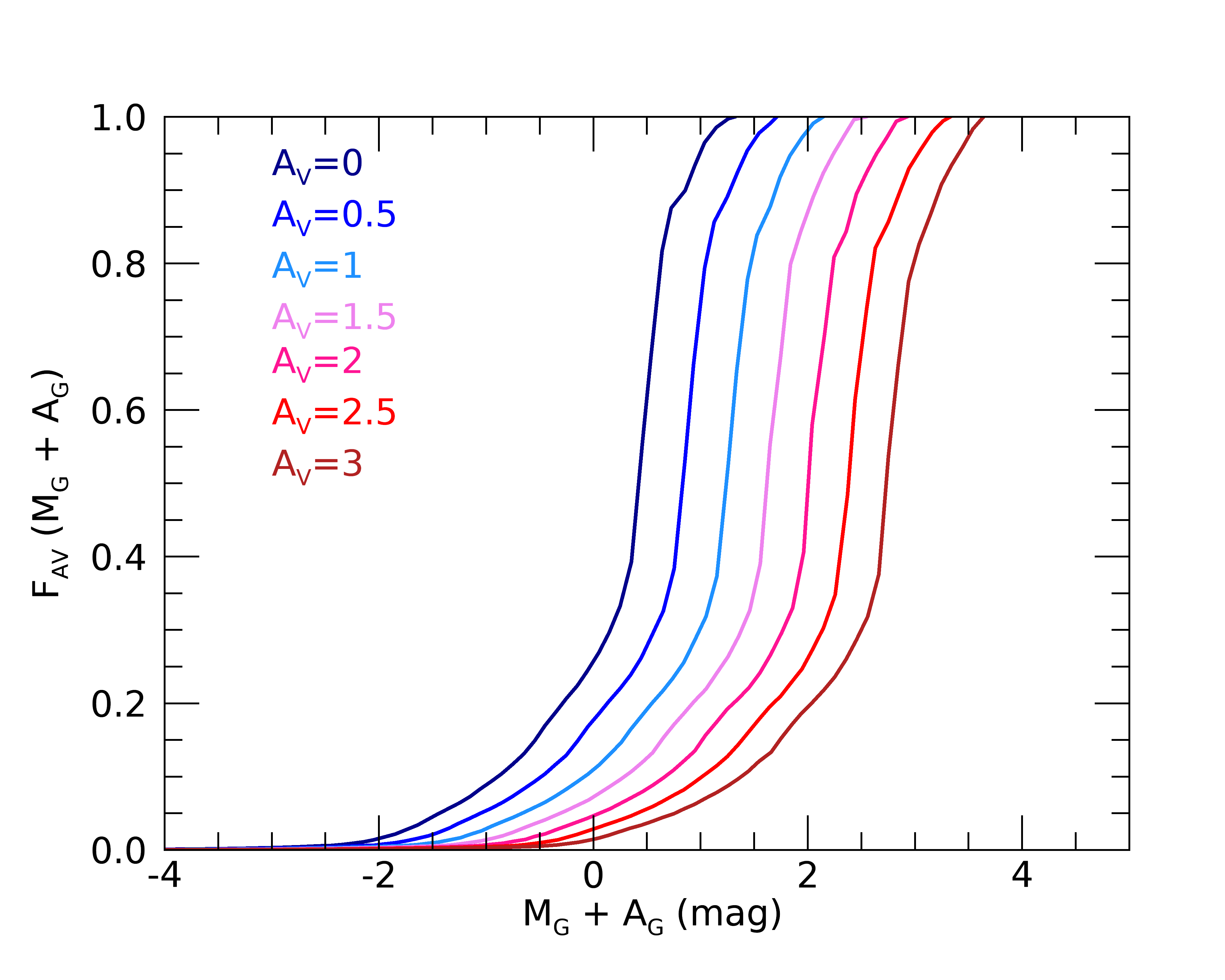} 
\caption{ The cumulative luminosity function for $\mathcal{M}_G=M_G + A_G$, obtained by applying to our modelled colour-magnitude diagram (see text) the color-color cuts performed to select our sample for different values of extinction. \label{Fig:lf}}
\end{centering}
\end{figure}

\begin{figure}[hb!]
\begin{centering} 
\includegraphics[trim={0.4cm 0.5cm 1cm 1cm},clip,width=0.42  \hsize]{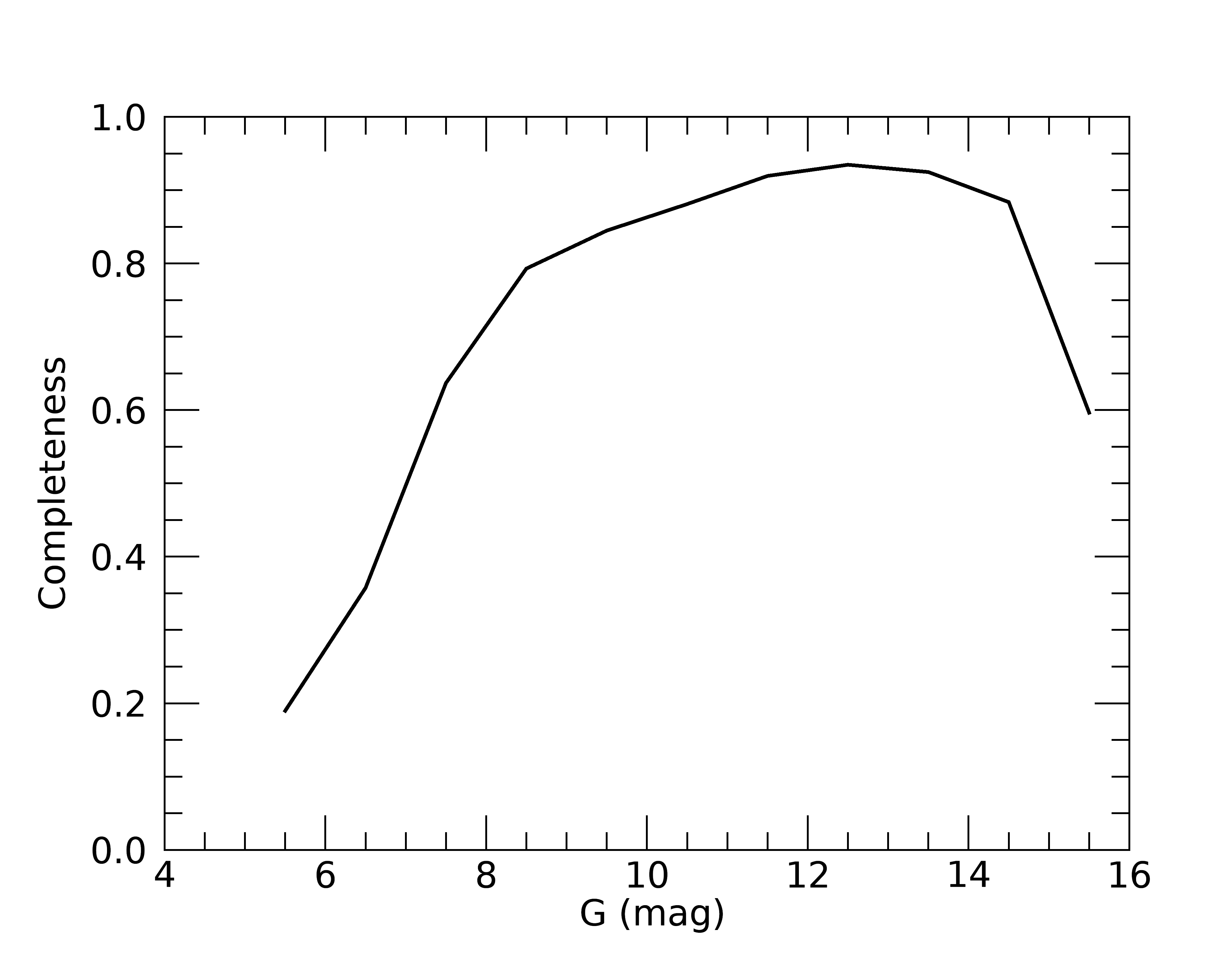} 
\caption{ The completeness of the dataset as a function of apparent magnitude G. \label{Fig:completeness}}
\end{centering}
\end{figure}

\newpage

\begin{figure}[hb!]
\begin{centering} 
\includegraphics[trim={0.cm 0.cm 0cm 0cm},clip,width=0.82  \hsize]{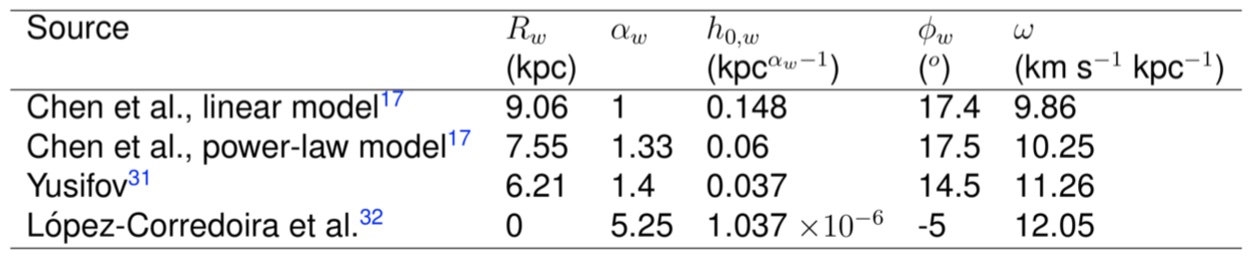}
\caption{ Shape parametrizations of the geometrical warp models adopted in this work, following Equation \ref{zwarp} and \ref{hzwarp}, and the obtained precession rates. The radius $R_w$ was scaled to account
for different assumptions about the Sun - Galactic center distance in this work and in the considered papers. \label{Tab:par_warp}}
\end{centering}
\end{figure}

\begin{figure}[hb!]
\begin{centering} 
\includegraphics[trim={0.cm 0.cm 0cm 0cm},clip,width=0.92  \hsize]{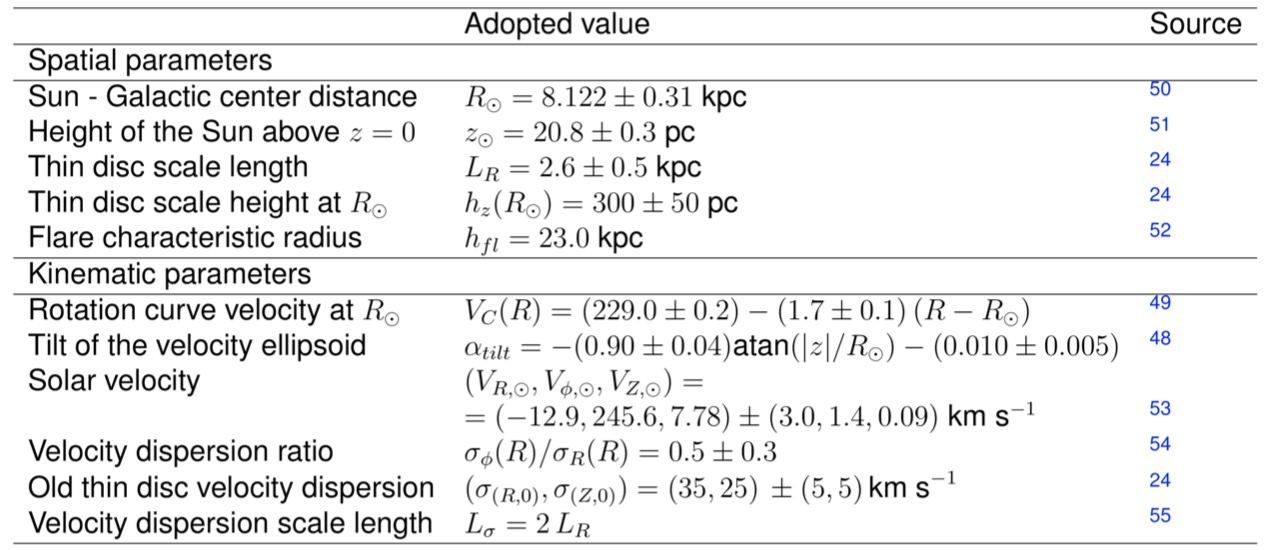} 
\caption{ Spatial and kinematic parameters adopted in this work. \label{Tab:par}}
\end{centering}
\end{figure}

\end{methods}




\newpage

\section*{References}

\bibliographystyle{naturemag}
\bibliography{abbrv_J,mybib2}


\begin{addendum}
 \item The authors thank the anonymous referees for their insightful comments that improved the quality of this manuscript. E. Poggio and R. Drimmel thank S. Casertano, V. Debattista and K. V. Johnston for useful discussions. E. Poggio thanks B. Bucciarelli for providing the software for the propagation of astrometric covariances. This work has made use of data from the European Space Agency (ESA) mission
{\it Gaia} (\url{https://www.cosmos.esa.int/gaia}), processed by the {\it Gaia}
Data Processing and Analysis Consortium (DPAC,
\url{https://www.cosmos.esa.int/web/gaia/dpac/consortium}). Funding for the DPAC
has been provided by national institutions, in particular the institutions
participating in the {\it Gaia} Multilateral Agreement. This work was supported by ASI (Italian Space Agency) under contract 2018-24-HH.0. This work was funded in part by the DLR (German space agency) via grant 50 QG 1403.
\item[Author contributions] EP contributed to the sample preparation, the modelling and data analysis, and wrote the manuscript together with RD. RD contributed to the model construction and interpretation of the results. RA, CBJ, MF helped with the statistical inference and revised the text. MGL, RLS and AS contributed to the project planning and revised the text.
 \item[Competing Interests] The authors declare that they have no competing financial interests.
 \item[Correspondence and request for materials] should be addressed to Eloisa Poggio~(email: eloisa.poggio@inaf.it). 
 \item[Data availability] The dataset can be downloaded at \url{https://figshare.com/articles/Giants_P18_csv/11382705}.
\end{addendum}


\end{document}